\documentclass[twocolumn,aps,prl,linenumber,showpacs,floatfix,superscriptaddress]{revtex4-1}
\usepackage{amsfonts}
\usepackage{amsmath}
\usepackage{amssymb}
\usepackage{dcolumn}
\usepackage{graphicx}
\usepackage{varwidth}
\usepackage{bm}
\usepackage{color}

\begin{document}

\newcommand{\HIM}{
Helmholtz Institute Mainz, D-55128 Mainz, Germany}

\newcommand{\NZIAS}{
Centre for Theoretical Chemistry and Physics,
The New Zealand Institute for Advanced Study,
Massey University Auckland, Private Bag 102904, 0632 Auckland, New Zealand}

\newcommand{\UNSW}{
School of Physics, University of New South Wales, Sydney 2052, Australia}

\newcommand{\MBU}{
Department of Chemistry, Faculty of Natural Sciences, Matej Bel University, 
Tajovsk\'{e}ho 40, SK-974 00 Bansk\'{a} Bystrica, Slovakia}

\newcommand{\CU}{
Department of Physical and Theoretical Chemistry, Faculty of Natural Sciences, Comenius University, Mlynsk\'a  dolina, 84104 Bratislava, Slovakia}

\title{Search for variation of fundamental constants: \\
Strong enhancements in $X^2\Pi$ cations of dihalogens and hydrogen halides}

\author{L. F. Pa\v steka}\email{lukas.f.pasteka@gmail.com}
\affiliation{\NZIAS}
\affiliation{\CU}

\author{A. Borschevsky}\email{a.borschevsky@massey.ac.nz}
\affiliation{\NZIAS}

\author{V. V. Flambaum}\email{v.flambaum@unsw.edu.au}
\affiliation{\UNSW}
\affiliation{\NZIAS}

\author{P. Schwerdtfeger}\email{p.a.schwerdtfeger@massey.ac.nz}
\affiliation{\NZIAS}

\date{\today}


\begin{abstract}
We propose to use diatomic molecular ions to search for strongly enhanced effects of variation of fundamental constants. The relative enhancement occurs in transitions between nearly degenerate levels of different nature. Since the trapping techniques for molecular ions have already been developed, 
the molecules HBr$^+$, HI$^+$, Br$_2^+$, I$_2^+$, IBr$^+$, ICl$^+$, and IF$^+$ are very promising candidates for such future studies.
\end{abstract}

\maketitle

Theories unifying gravity with the other interactions suggest the possibility of spatial and temporal variation of dimensionless fundamental physical constants, such as the fine structure constant, $\alpha=e^2/\hbar c$ and the proton to electron mass ratio, $\mu = m_p/m_e$ \cite{Bar05,Uza11}, the latter being connected to the strong coupling constant \cite{FlaShu02,FlaShu03}.
Two major directions for the search of $\alpha$ and $\mu$ variation are observational studies, such as the analyses of high-resolution quasar absorption spectra (Refs. \cite{DzuFlaWeb99,LevLapHen10,WebKinMur11} and references therein), and laboratory research, including atomic and molecular clocks \cite{FlaDzu09,ChiFlaKoz09}.

Molecules are very promising probes for variation of fundamental constants (VFC) as molecular spectra can be very sensitive to both $\alpha$ and $\mu$ \cite{ChiFlaKoz09,KozLev13,JanBetUba14}, making it possible to look for variation of both constants in a single experiment. So far, the use of molecules has mostly focused on astrophysical observations of molecular spectra at high red-shifts.  Molecular probes used in astrophysical research include the Lyman and Werner absorption lines in H$_2$ and HD molecules  \cite{MalBunMur10,WeeMurMal11}, the 18 cm$^{-1}$ transition of OH \cite{KanCarLan05}, the inversion spectrum of ammonia \cite{FlaKoz07_2,MurFlaMul08,HenMenMur09}, $\Lambda-$doubling transitions of CH \cite{TruTokLew13},  and the internal rotational transitions of interstellar methanol \cite{BagJanHen13}. 

The most stringent laboratory molecular limit to date on the variation of $\mu$ is $\dot \mu /\mu =(-3.8 \pm 5.6)\times 10^{-14}$ yr$^{-1}$ \cite{SheButCha08}, set in a two-year-long experiment over the course of which vibrational transitions in the SF$_6$ molecule were compared with the Cs standard. So far, this is the only laboratory experiment using molecular systems to study VFC. However, the development of methods for creation and trapping of ultracold molecules \cite{DanHalGus08,NiOspMir08} opens exciting new prospects to test for VFC by means of high-resolution molecular spectroscopy.
In such experiments it is important to use molecules that experience enhanced sensitivity to VFC. However, locating such favourable systems is not a trivial task.

Nearly degenerate levels with different sensitivity to VFC may provide huge enhancements of a relative variation, since $\delta\omega/\omega$ tends to infinity when the distance between the levels $\omega$ is close to zero. This is the case with dysprosium \cite{DzuFlaWeb99,DzuFlaWeb99_2}. Relatively low accuracy of the transition frequency measurements in Dy provides record-high accuracy in VFC measurements \cite{LeeWebCin13}. However, the problem with Dy is that one of the levels has a short lifetime, thus limiting the accuracy.

Molecules can have both degenerate levels and very narrow transition widths. Diatomic molecules that have a near cancellation between hyperfine structure and rotational intervals \cite{Fla06_2} or between fine structure and vibrational intervals \cite{FlaKoz07} can benefit from such an enhanced sensitivity. A number of such molecules have already been identified, e.g. Cs$_2$ \cite{DemSaiSag08,BelBorFla11}, CaH, MgH, CaH$^+$ \cite{Kaj08,KajMor09,AbeMorHad12}, Cl$_2^+$, IrC, HfF$^+$, SiBr, LaS, LuO, and others \cite{Fla06_2,FlaKoz07,BelBorSch10}. Several polyatomic molecules were suggested for laboratory studies of VFC as well, see e.g. \cite{QuiWalHoe14,JanKleMen13} and references in \cite{JanBetUba14,ChiFlaKoz09}.
Here we note that practical advantage can be gained by using molecular ions rather than neutral molecules due to the greater ease of trapping and cooling \cite{MeyBohDes06,NguVitHoh11}. Unfortunately, experimental spectroscopic constants for charged diatomics are scarce compared to their neutral counterparts, 
and the search for VFC using such molecules is somewhat limited.

In this work we investigate a group of diatomic molecules that satisfy all of the above mentioned criteria, namely the singly charged dihalogens and hydrogen halides. We are interested in the 
forbidden rovibrational transitions between the nearly degenerate sub-levels of the $X^2 \Pi$ ground states of these molecules. Using available experimental spectroscopic data, we identify such promising transitions, and examine the dependence of the transition energy on both $\alpha$ and $\mu$. Strong enhancement of the sensitivity of these transitions to variations in $\alpha$ and in $\mu$ makes them excellent candidates for future experiments to detect VFC.

The systems presented here were selected for an in-depth study due to the availability of their accurate experimental spectroscopic parameters, which are listed (along with their definitions) in the Supplementary Material at [URL will be inserted by publisher]. These systems include four isotopologues of HBr$^+$  (H$^{79}$Br$^+$ , H$^{81}$Br$^+$, D$^{79}$Br$^+$, and D$^{81}$Br$^+$) \cite{ZeiBohNel87,ChaHoDal95}, two isotopologues of HI$^+$ (HI$^+$ and DI$^+$) \cite{ChaHo95}, three isotopologues of  Br$_2^+$ ($^{79}$Br$_2^+$, $^{81}$Br$_2^+$, and $^{79}$Br$^{81}$Br$^+$) \cite{HarElaTuc83}, I$_2^+$ \cite{CocDonLaw96,DenLunZhu12}, two isotopologues of IBr$^+$ (I$^{79}$Br$^+$ and I$^{81}$Br$^+$) \cite{BeaMacKen98}, two isotopologues of ICl$^+$ (I$^{35}$Cl$^+$ and I$^{37}$Cl$^+$) \cite{YenLopKin00}, and IF$^+$ \cite{InnFabEyp08}. 
A number of other charged dihalogens and hydrogen halides were considered, but were found unsuitable for the present study. These systems are discussed in more detail in the Supplementary Material.

Using available spectroscopic constants, we reproduce the molecular potential energy curves and the spectroscopic levels by following the Rydberg-Klein-Rees procedure \cite{Ryd32,Kle32,Ree47,Sen95}. This allows us to locate the promising low-energy rovibrational transitions between the two substates of the $X ^2 \Pi$ ground state of the molecules of interest. Such low energy transitions will exhibit a strong enhancement of sensitivity to VFC.
Once the favourable transitions are located, we examine the dependence of their energy on the $\alpha$ and $\mu$ constants. Following the line of thought from  earlier work \cite{FlaKoz07,BelBorSch10} we derive simple analytical expressions for the transition energies and their fractional variation. Here we do not deal  with the lowest vibrational and rotational levels only, but also include the effect of  anharmonicity and rotational spacings, which was neglected in earlier publications. Taking only the harmonic frequency $\omega_e$, the first anharmonicity $\omega_e x_e$, the rotational constant $B_e$, and the spin-orbit coupling constant $A_e$ and its first vibrational dependence $A^{(1)}$ into account (thus neglecting higher order terms), the transition energy $\omega$ between the two states can be expressed as
\begin{align}
\label{eq:trans} 
\omega= E'-E''=&\ \omega_e(\nu'-\nu'') \nonumber \\
&-\omega_e x_e(\nu'-\nu'')(\nu''+\nu'+1) \nonumber \\ 
 &+B_e(J'-J'')(J''+J'+1) \\
  &- A_e - \frac{1}{2}A^{(1)}(\nu''+\nu'+1), \nonumber
\end{align}
where we denote the initial state by double prime and the final state by single prime; in this case these correspond to the $X^2\Pi_{\frac{3}{2}}$ and the $X^2\Pi_{\frac{1}{2}}$ states, respectively.

The spin-orbit coupling constant \(A_v\) and its constituents $A_e$, $A^{(1)}$, $A^{(2)}$, etc. scale as \(\sim\alpha^2\). The rest of the spectroscopic constants are only weakly dependent on \(\alpha\) and this dependence can be neglected \cite{BelBorSch10}. Following from the dependence of various spectroscopic constants on the reduced mass \cite{Mul25}, we can easily deduce their scaling with the proton to electron mass ratio \(\mu\): \(\omega_e\) and \(A^{(1)}\) scale as \(\sim\mu^{-\frac{1}{2}}\), while  \(\omega_e x_e\) and \(B_e\) scale as \(\sim\mu^{-1}\).

Using these analytical dependencies along with Eq. \ref{eq:trans}, the fractional variation of the transition energy $\omega$ can be calculated as
\begin{equation}
\frac{\delta\omega}{\omega}=K_{\mu}\frac{\delta\mu}{\mu}+K_{\alpha}\frac{\delta\alpha}{\alpha},
\end{equation}
with the relative enhancement factors \(K_{\mu}\) and \(K_{\alpha}\) defined as
\begin{align}\label{eq:Kmu}
K_{\mu}=&[\frac{1}{2}\omega_e(\nu''-\nu') - \omega_e x_e(\nu''-\nu')(\nu''+\nu'+1) \nonumber \\
 &+ B_e(J''-J')(J''+J'+1)  \\
 &+\frac{1}{4}A^{(1)}(\nu''+\nu'+1)]\omega^{-1}= \tilde{K}_{\mu}\omega^{-1}, \nonumber
\end{align}
\begin{equation}\label{eq:Kalpha}
K_{\alpha}=[- 2A_e - A^{(1)}(\nu''+\nu'+1)]\omega^{-1} = \tilde{K}_\alpha\omega^{-1},
\end{equation}
where the symbols $\tilde{K}_{\alpha}$ and $\tilde{K}_\mu$ represent the absolute enhancement factors.

\begin{figure}
  \centering
  \includegraphics[width=1.0\linewidth]{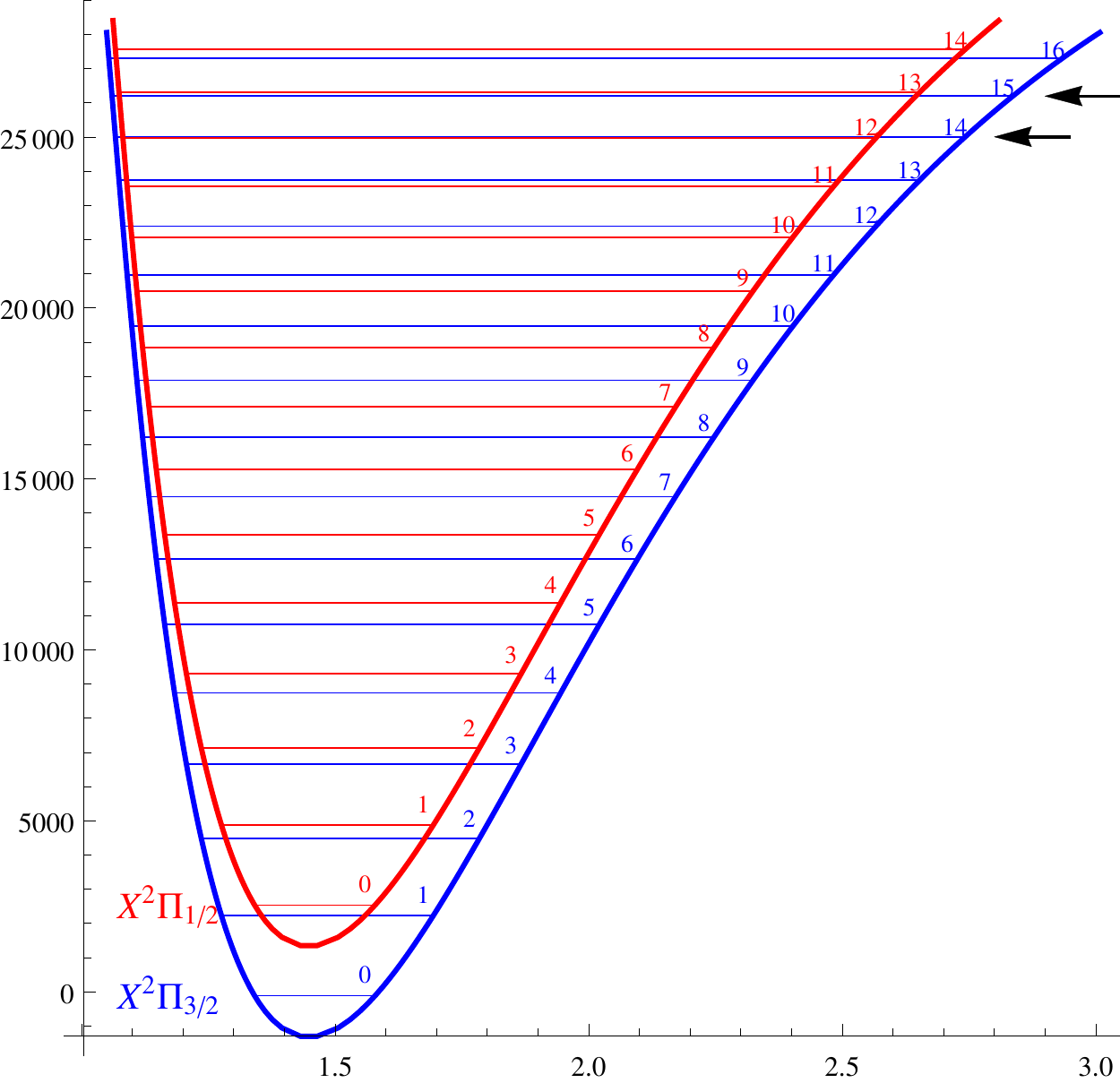}
  \caption{(color online) Potential energy curves of the two substates of the $X ^2 \Pi$ state in H$^{79}$Br$^+$. Arrows point to the quasi-degenerate vibrational levels. Origin of the vertical axis corresponds to the bottom of the unsplit potential well.}
  \label{fig:vib}
\end{figure}

We have examined the effect of including the next order parameters in the Dunham series, i.e. $\omega_e y_e$, $\alpha_e$, and $A^{(2)}$, on the calculated \(K_{\mu}\) and \(K_{\alpha}\), and found it to be negligible for the majority of the cases considered here.

\begin{figure}
  \begin{tabular}{ll}
\raisebox{-.9\height}{\includegraphics[scale=0.66]{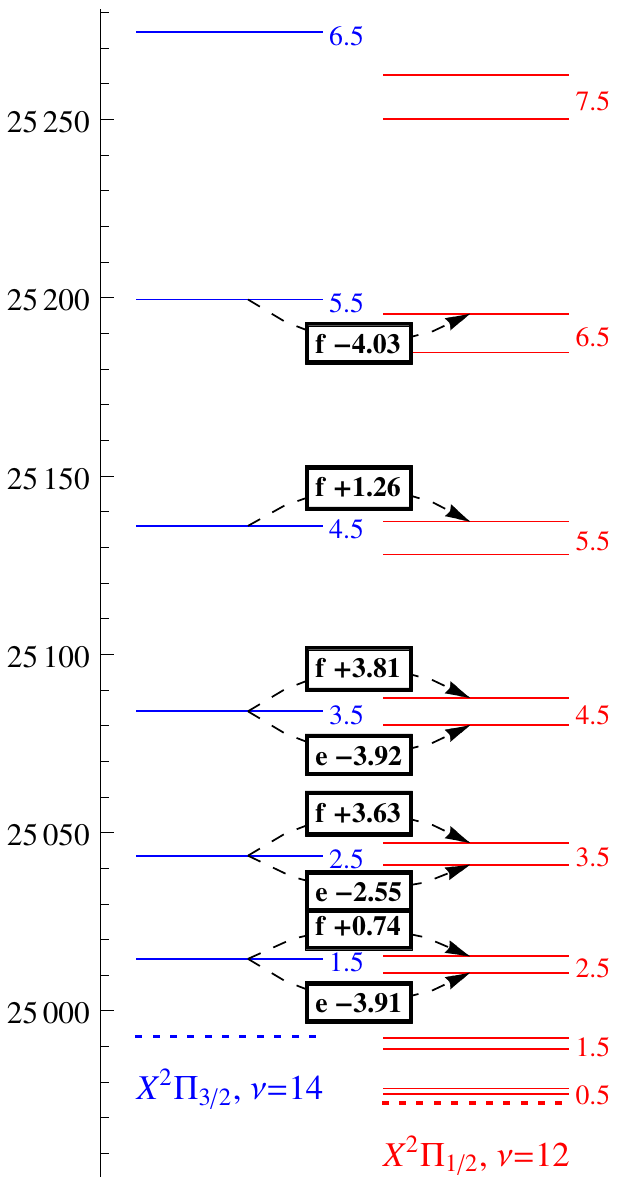}} & \raisebox{-.9\height}{\includegraphics[scale=0.66]{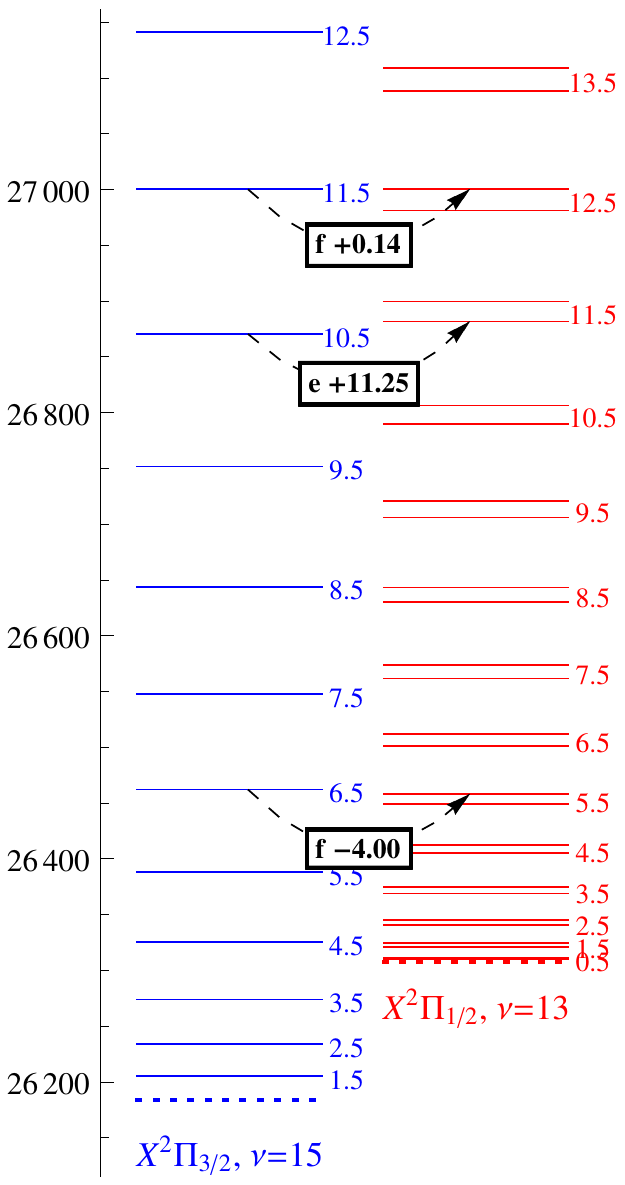}} 
  \end{tabular}
  \caption{(color online) Rotational structure of the selected quasi-degenerate vibrational levels of H$^{79}$Br$^+$ with transitions represented by dashed arrows. Allowed $e \leftrightarrow e$ and $f \leftrightarrow f$ transitions are denoted with $e$ and $f$, respectively, in the brackets with transition energies (cm$^{-1}$).}
  \label{fig:rot}
\end{figure}

We illustrate the scheme outlined above by using H$^{79}$Br$^+$ as an example. Figure \ref{fig:vib} presents the reconstructed potential energy curves of the two sub-states of the $X ^2 \Pi$ ground state of H$^{79}$Br$^+$. We find that the $X ^2 \Pi_\frac{3}{2}$, $\nu=14$ level is very close in energy to the $X ^2 \Pi_\frac{1}{2}$, $\nu=12$ level, differing by 19 cm$^{-1}$.
We then take a closer look at the rotational states of the favourable vibrational transition, in order to select even closer quasi-degeneracies of the energy levels (Figure \ref{fig:rot}, left). Vibrational levels $X ^2 \Pi_\frac{3}{2}$, $\nu=15$ and $X ^2 \Pi_\frac{1}{2}$, $\nu=13$, although futher apart (123 cm$^{-1}$), also exhibit quasi-degeneracies in the rotational structure (Figure \ref{fig:rot}, right).

We focus only on the dominant single-photon dipole transitions obeying the $\Delta J=0, \pm1$ selection rule. Since the transitions conserving the total angular momentum ($\Delta J=0$) can be blended together for all values of $J$ \cite{BelBorSch10}, thus negatively affecting the accuracy, we are interested in the $\Delta J=\pm1$ transitions only. Furthermore, we also include the effect of $\Lambda$-doubling and we only consider transitions following the rotational symmetry selection rule $e \nleftrightarrow f$. The $\Lambda$-doubling of the $^2\Pi_\frac{3}{2}$ rotational levels is negligible and not even visible in the scope of Figure \ref{fig:rot}.
We can identify a number of rovibrational transitions corresponding to a unit change in the total angular momentum $J$ while having very low transition energies between the reconstructed rovibrational levels. These are listed in Table \ref{tab:enhanc}, along with the transition energies and the corresponding enhancement factors as defined in Eqs. \ref{eq:Kmu} and \ref{eq:Kalpha}.  The same procedure is performed for the rest of the systems under study. These results are also listed in Table \ref{tab:enhanc}, and the corresponding plots are presented in the Supplementary Material. Note that there are more favorable transitions identified in the rotational plots, as these usually occur in series for several neighbouring rotational levels, however, we only present the ones with lowest $\omega$ (and therefore highest $K_\mu$ and $K_\alpha$) in Table \ref{tab:enhanc}.

\begin{table}
\setlength{\tabcolsep}{0.06cm}
\caption{Transition energies $\omega$ (cm$^{-1}$) and corresponding enhancement factors, absolute, $\tilde{K}_\mu$ and $\tilde{K}_\alpha$ (cm$^{-1}$), and relative, $K_\mu$ and $K_\alpha$, for several identified transitions of cations under investigation. Symbols $e$ and $f$ denote the transitions of type $e \leftrightarrow e$ and $f \leftrightarrow f$, respectively.}
\footnotesize
\begin{tabular}{lrrrrrrrrrr}
\hline
cation & $\nu''$ & $\nu'$ & $J''$ & $J'$ & $s$ & $\omega$ & $\tilde{K}_\mu$ & $K_\mu$ & $\tilde{K}_\alpha$ & $K_\alpha$ \\
\hline
H$^{79}$Br$^+$	&	14	&	12	&	1.5	&	2.5	&	$f$	&	0.74	&	--80.8	&	--110	&	5250	&	7120 \\
&&&	4.5	&	5.5	&	$f$	&	1.26	&	--129	&	--103	&		&	4180 \\
&	15	&	13	&	11.5	&	12.5	&	$f$	&	0.14	&	--426	&	--3030	&	5250	&	37200 \\
H$^{81}$Br$^+$	&	14	&	12	&	1.5	&	2.5	&	$f$	&	0.82	&	--80.4	&	--97.9	&	5250	&	6390 \\
&&&	5.5	&	6.5	&	$f$	&	--0.27	&	--145	&	543	&		&	--19700 \\
&	15	&	13	&	11.5	&	12.5	&	$e$	&	--1.44	&	--426	&	295	&	5250	&	--3630 \\
&&&	6.5	&	5.5	&	$f$	&	0.52	&	--119	&	--228	&		&	10100 \\
D$^{79}$Br$^+$	&	9	&	7	&	7.5	&	8.5	&	$f$	&	2.20	&	881	&	400	&	5270	&	2390 \\
&	10	&	8	&	1.5	&	0.5	&	$e$	&	1.99	&	871	&	437	&	5270	&	2650 \\
D$^{81}$Br$^+$	&	9	&	7	&	7.5	&	8.5	&	$f$	&	2.49	&	881	&	354	&	5270	&	2120 \\
&	10	&	8	&	1.5	&	0.5	&	$e$	&	2.23	&	871	&	391	&	5270	&	2370 \\
HI$^+$	&	6	&	3	&	3.5	&	4.5	&	$e$	&	--3.05	&	2060	&	--675	&	10700	&	--3490 \\
&&&	21.5	&	20.5	&	$f$	&	--0.41	&	2390	&	--5780	&		&	--25800 \\
&&&	26.5	&	25.5	&	$e$	&	2.63	&	2450	&	934	&		&	4060 \\
&	12	&	8	&	7.5	&	8.5	&	$f$	&	1.50	&	984	&	656	&	10600	&	7070 \\
&&&	21.5	&	20.5	&	$e$	&	--2.35	&	1360	&	--580	&		&	--4510 \\
&	16	&	11	&	16.5	&	15.5	&	$e$	&	0.12	&	190	&	1630	&	10600	&	91000 \\
DI$^+$	&	7	&	3	&	1.5	&	2.5	&	$e$	&	0.27	&	2230	&	8220	&	10700	&	39300 \\
&&&	21.5	&	20.5	&	$f$	&	0.98	&	2390	&	2420	&		&	10800 \\
&&&	27.5	&	26.5	&	$e$	&	2.42	&	2430	&	1000	&		&	4420 \\
&	14	&	9	&	9.5	&	10.5	&	$e$	&	--0.91	&	1460	&	--1600	&	10600	&	--11700 \\
&&&	22.5	&	21.5	&	$f$	&	--0.76	&	1670	&	--2190	&		&	--14000 \\
&&&	25.5	&	24.5	&	$e$	&	--3.92	&	1690	&	--430	&		&	--2710 \\
&	19	&	3	&	9.5	&	10.5	&	$f$	&	1.08	&	686	&	638	&	10600	&	9860 \\
&&&	19.5	&	18.5	&	$f$	&	--3.64	&	878	&	--241	&		&	--2910 \\
I$_2^+$	&	30	&	6	&	30.5	&	31.5	&	$e$	&	--0.00	&	2270	&	--682000	&	9990	&	--3010000 \\
&&&	34.5	&	35.5	&	$f$	&	--0.06	&	2260	&	--40000	&		&	--176000 \\
&	39	&	14	&	50.5	&	51.5	&	$e$	&	0.01	&	2090	&	230000	&	9800	&	1080000 \\
&&&	54.5	&	55.5	&	$f$	&	--0.13	&	2090	&	--16200	&		&	--76200 \\
&	48	&	22	&	1.5	&	2.5	&	$f$	&	0.06	&	1900	&	30500	&	9610	&	155000 \\
&&&	19.5	&	18.5	&	$e$	&	--0.02	&	1900	&	--79100	&		&	--401000 \\
&&&	28.5	&	27.5	&	$f$	&	--0.04	&	1900	&	--45900	&		&	--232000 \\
$^{79}$Br2$^+$	&	11	&	3	&	22.5	&	23.5	&	$e$	&	0.15	&	1350	&	9170	&	5570	&	37900 \\
&&&	24.5	&	25.5	&	$f$	&	--0.10	&	1350	&	--13700	&		&	--56600 \\
&	31	&	22	&	18.5	&	19.5	&	$e$	&	--0.10	&	1190	&	--12300	&	5370	&	--55400 \\
&&&	21.5	&	22.5	&	$f$	&	0.04	&	1190	&	33200	&		&	150000 \\
$^{79,81}$Br$_2^+$	&	9	&	1	&	20.5	&	21.5	&	$e$	&	0.09	&	1360	&	14500	&	5600	&	59700 \\
&&&	22.5	&	23.5	&	$f$	&	--0.11	&	1360	&	--12700	&		&	--52200 \\
&	30	&	21	&	3.5	&	4.5	&	$e$	&	--0.09	&	1160	&	--12900	&	5450	&	--60400 \\
&&&	4.5	&	5.5	&	$f$	&	--0.07	&	1160	&	--15500	&		&	--73000 \\
$^{81}$Br$_2^+$	&	8	&	0	&	12.5	&	13.5	&	$e$	&	0.11	&	1370	&	12900	&	5600	&	52700 \\
&&&	14.5	&	15.5	&	$f$	&	0.09	&	1370	&	14500	&		&	59400 \\
&	28	&	19	&	36.5	&	37.5	&	$e$	&	0.05	&	1190	&	25700	&	5450	&	118000 \\
&&&	39.5	&	40.5	&	$f$	&	--0.10	&	1190	&	--11600	&		&	--53400 \\
I$^{79}$Br$^+$	&	25	&	8	&	27.5	&	28.5	&	$e$	&	--0.05	&	1720	&	--32200	&	9220	&	--173000 \\
&&&	30.5	&	31.5	&	$f$	&	0.20	&	1720	&	8440	&		&	45300 \\
&	39	&	22	&	28.5	&	29.5	&	$e$	&	0.04	&	1020	&	25200	&	9140	&	225000 \\
&&&	30.5	&	31.5	&	$f$	&	0.02	&	1020	&	63400	&		&	566000 \\
I$^{81}$Br$^+$	&	24	&	7	&	31.5	&	32.5	&	$e$	&	0.06	&	1760	&	31700	&	9320	&	168000 \\
&&&	34.5	&	35.5	&	$f$	&	0.23	&	1760	&	7630	&		&	40500 \\
&	36	&	19	&	7.5	&	6.5	&	$f$	&	--0.01	&	1170	&	--83400	&	9320	&	--664000 \\
&&&	12.5	&	11.5	&	$f$	&	--0.01	&	1170	&	--88700	&		&	--707000 \\
I$^{35}$Cl$^+$	&	20	&	8	&	18.5	&	19.5	&	$e$	&	0.31	&	1940	&	6160	&	9550	&	30300 \\
&&&	20.5	&	21.5	&	$f$	&	--0.01	&	1940	&	--166000	&		&	--817000 \\
I$^{37}$Cl$^+$	&	20	&	8	&	18.5	&	19.5	&	$e$	&	--0.11	&	1940	&	--17700	&	9550	&	--87000 \\
&&&	21.5	&	22.5	&	$f$	&	0.15	&	1940	&	12800	&		&	62900 \\
IF$^+$	&	14	&	5	&	23.5	&	24.5	&	$e$	&	--0.06	&	2760	&	--45900	&	11200	&	--186000 \\
&&&	26.5	&	27.5	&	$f$	&	--0.12	&	2760	&	--23500	&		&	--95200 \\
&	22	&	12	&	34.5	&	35.5	&	$e$	&	0.49	&	2780	&	5680	&	11100	&	22600 \\
&&&	37.5	&	38.5	&	$f$	&	0.30	&	2780	&	9260	&		&	36900 \\
&	28	&	17	&	34.5	&	35.5	&	$e$	&	--0.66	&	2820	&	--4270	&	11000	&	--16600 \\
&&&	37.5	&	38.5	&	$f$	&	--0.41	&	2820	&	--6820	&		&	--26500 \\
\hline
\end{tabular}
\label{tab:enhanc}
\end{table}

The obtained values of $K_{\alpha}$ look very promising and span orders of magnitude ranging from $10^3$ to $10^6$. Several transitions in I$_2^+$, I$^{81}$Br$^+$, and I$^{35}$Cl$^+$ are particularly sensitive to variation in $\alpha$, with $K_{\alpha}$ values larger than 600,000, due to the almost perfect degeneracy between these levels, differing by only about $0.01$ cm$^{-1}$.

The calculated $K_{\mu}$ are generally one or two orders of magnitude smaller than $K_{\alpha}$. However, many theories suggest that the  variation in $\mu$ should be on the order of ~30 times larger than that in $\alpha$ (see e.g. \cite{FlaLeiTho04} and references therein). Thus, the effect of possible $\mu$-variation on the spectra of these systems should not be neglected.
Since the first term in Eq. \ref{eq:Kmu} is dominant, we can easily see, that the larger the difference between initial and final state vibrational quantum number $\nu$, the higher is the enhancement factor $K_\mu$. This is automatically satisfied for heavy diatomics with larger spin-orbit splitting between the two $X^2\Pi$ sub-states as is illustrated by the results in Table \ref{tab:enhanc}.

The identified transitions in the HBr$^+$ and HI$^+$ isotopologues, as well as in $^{79}$Br$^{81}$Br$^+$, may be of particular interest for experimental investigation because of their low-$J$ nature. Rotationally cooled molecules should produce spectra with higher resolution \cite{VogMadDre06,StaHojKla10,SheBorHan12,HanVerKlo14}, which is of great importance for detection of the minute changes in transition energies resulting from VFC.
    
An advantage can be gained by considering the variation of the ratio of two transition frequencies within the same rovibronic spectrum, $\omega_1/ \omega_2$, rather than the variation of the frequencies themselves \cite{BelBorSch10}. The variation of $\omega_1/ \omega_2$ is given by

\begin{equation}
\frac{\delta(\omega_1/ \omega_2)}{(\omega_1/ \omega_2)}=(K_{\mu}^1-K_{\mu}^2) \frac{\delta\mu}{\mu}+(K_{\alpha}^1-K_{\alpha}^2) \frac{\delta\alpha}{\alpha},
\end{equation}

Considering the variation of a dimensionless  ratio of two transition energies removes the dependence on the unit system and minimizes the systematic effects. In addition, selecting two transitions where either $K_{\mu}^1$ and $K_{\mu}^2$,  or $K_{\alpha}^1$ and $K_{\alpha}^2$, or both pairs, have an opposite sign, maximizes the sensitivity of the measurement.

\textbf{Acknowledgements:}
Funding through the Marsden Fund (MAU0606) and Australian Research Coucil is gratefully acknowledged. A.B. is grateful to S. Lemmens for fruitful discussions.

\bibliography{2Pi_systems}

\end{document}